\documentstyle[editedvolume,numreferences]{crckapb}

\begin{opening}
\title{GENERALIZED HUBBLE LAW, \protect\\ VIOLATION OF THE COSMOLOGICAL PRINCIPLE AND \protect\\ SUPERNOVAE }
\author{J.-F. PASCUAL-SANCHEZ}
\institute{Universidad de Valladolid\\
            Dept. Matem\'atica Aplicada Fundamental,\\ Secci\'on
Facultad de Ciencias, 47005, Valladolid, Spain}
\end{opening}

\runningtitle{GENERALIZED HUBBLE LAW}

\begin{document}

% The \begin{document} command comes after the \end{opening}
% command.
\begin{abstract} { The acceleration of the cosmic
expansion has been discovered  as a consequence of redshift
Supernovae data. In the usual way, this cosmic acceleration is
explained by the presence of a positive cosmological constant or
quantum vacuum energy, in the background of standard Friedmann
models. Recently, looking for an alternative explanation, I have
considered an inhomogeneous barotropic spherically symmetric
spacetime. Obviously, in this inhomogeneous model the
philosophical cosmological principle is not verified. Within this
framework, the kinematical acceleration of the cosmic fluid or,
equivalently, the inhomogeneity of matter, is just the responsible
of the SNe Ia measured cosmic acceleration. Moreover, this model
gives rise to a generalized Hubble law with two anisotropic terms
(dipole acceleration and quadrupole shear), besides the expansion
one. The dipole term of this generalized Hubble law could explain,
in a cosmological setting, the observed large scale flow of
matter, without to have recourse to peculiar velocity-type
newtonian models which assume a Doppler dipole.}
\end{abstract}

\section{Introduction}
Recently [1], I have considered a specific inhomogeneous
 cosmological model, which could explain, in an alternative way,
 by the presence of
 a  kinematic acceleration generated by a
 negative gradient of pressure (or mass-energy), the present negative
  deceleration parameter, which appears to be a result of
 high redshift Supernovae data [2]. In the current way, these data are explained
 by
  the presence of a positive cosmological constant or vacuum
  energy or quintessence, in the
   background of standard Friedmann (FLRW)
  models with perfect fluid matter.

In this specific inhomogeneous model, which I will also consider
 in this work, the cosmic matter is described by a barotropic (B) perfect fluid and
 geometrically has
  spherical (S)
  symmetry (hereafter BS). The comoving perfect fluid
matter congruence, in this vorticity-free BS inhomogeneous model,
has expansion, kinematic acceleration and shear, at difference
with the standard FLRW models, where it has expansion only.

\section{The BS model can explain cosmic acceleration}
  In the BS model, the isotropy group
is 1-dim, whereas the isometry group is a 3-dim $G_3$, which is
acting multiply transitively on spacelike 2-dim surfaces
orthogonal to a preferred direction (hereafter $e_1$). These 2-dim
surfaces orbits, orthogonal both to the 4-velocity  $u^a$ of the
congruence and to $e_1$, are spheres.

 Also, in the BS model
 the vorticity is zero. This implies both the
existence of a global cosmic time and of 3-dim global spacelike
hypersurfaces orthogonal to the fluid congruence.

 Thus, the metric in comoving coordinates has the
expression
\begin{equation}
ds^2=-N^2(r,t)\, dt^2+B^2(r,t)\, dr^2+R^2(r,t)\, d\Omega^2 .
\end{equation}
where $d\Omega^2$ is the spherical line element and the
coefficient $N(r,t)$ is a lapse function which relates global
cosmic and local proper times.
 In the BS spacetime exists a preferred central worldline,
 i.e.
  the spacelike hypersurfaces have a
 centre at $r = 0$, where the isotropy group is 3-dim,
 and a preferred radial direction $e_1$ at each point,
associated with the direction of the only non-null component, $A$,
of the kinematic acceleration of the matter fluid elements. This
kinematic acceleration satisfies the spatially contracted
equations of motion
\begin{equation}\label{4}
 p^{\prime}+ (\mu +p)~A=0,
\end{equation}
where a prime denotes the derivative along the preferred radial
direction with unit vector field $e_1$. We see from the last
formula that the kinematic acceleration, which opposes to the
gravitational attraction towards the centre, is generated by a
negative gradient of pressure or, equivalently, due to the
supposed barotropic equation of state $p(r,t) = p (\mu (r,t))$, by
a negative gradient of mass-energy density $\mu(r,t)$.

 Note (see
[1]), that the presence in the metric of the coefficient $N(r,t)$,
gives rise to a non null kinematic acceleration and also to a new
term in the expression of the deceleration parameter $q$,
\begin {equation}\label{14}
q_0 = \frac{1}{2}\Omega_0- I\!\!I_0,
\end{equation}
where $\Omega_0$ is the present matter density in units of the
critical density and $I\!\!I_0$  was called in [1] the
inhomogeneity parameter.

This additional inhomogeneity term $I\!\!I_0$ is positive and
hence the deceleration parameter $q_0$ can be negative at present
cosmic time, so one has an alternative explanation for the cosmic
acceleration implied by the Supernovae data, in the realm of the
BS model.

\section{Cosmological principle}
 As the
Cosmological Principle (CP) is not verified in the BS model, we
would like to remind that nowadays there are not observational
proofs of this philosophical assumption. The observational almost
isotropy of the cosmic background radiation (CBR) temperature is
insufficient
       to force exact isotropy into the spacetime geometry and hence exact spatial
        homogeneity of the 3-dim cosmic hypersurfaces, i.e., to force
        the verification of the Cosmological Principle, because neither the "exact"
Ehlers-Geren-Sachs (EGS) theorem nor the "almost EGS theorem" [3]
are necessarily verified.

 Thus the measured almost isotropy
of the CBR temperature is in principle compatible with large shear
[4] and (or) nonzero kinematic acceleration, as happens in the BS
model.

\section{Linear Hubble law for the BS model}
Now, we specialize the general linear Hubble law [5], valid for
any inhomogeneous universe, for the specific case of the BS model
considered. The final expression  that adopts  for off-center
observers $P_0$, is [6]:
\begin{equation}\label{8}
    z = \left( \frac{\dot{R}}{R}-A\cos \Psi+
\sqrt{3}\,\sigma\cos ^2 \Psi\right)_0\, D
\end{equation}
By spherical symmetry, just the telescopic angle $\Psi$ is needed
to describe off-centre observations in the rest space of the
observer. $\Psi$ is the angle between the direction of observation
of a light ray and the preferred vector $e_1$, $\sigma$ is the
scalar shear and $D$ is any cosmological distance.
 The generalized
Hubble function has three terms which contribute to the
cosmological redshift $z$. The first is similar to the usual one
of FLRW models due to the volume expansion, but in this model is
due to the azimuthal expansion and depends not only on cosmic time
but also on a radial comoving distance or position of the observer
with respect to the centre. The second term in the generalized
Hubble function is a dipolar one, due to the acceleration, and the
third is a quadrupolar one, due to the shear.

The consequences of the generalized linear Hubble law are
striking. For instance, when we are observing in the same sense
that the acceleration away from the centre, that is $\Psi=0$, then
the dipole term gives a maximum violetshift contribution, $-AD$.
Of course, observing in the opposite sense to the acceleration,
i.e., towards the centre, it gives an maximum dipole redshift,
$+AD$. In both cases, if the emitters are at the same distance,
the additional expansion and shear quadrupole terms have the same
positive value. Hence, using the new Hubble law (\ref{8}), the
difference of redshifts of these kind of observations, is a pure
dipole violetshift. Therefore, we consider this specific direction
away from the centre, as the global direction of the matter
dipole.

\section{Final comments and conclusions}
The usual dipole, manifested as a deviation of the isotropic
Hubble law of Friedmann models, is  a constant Doppler effect.
This Doppler effect arises in the standard model from peculiar
velocities and it can be eliminated going to the correct zero
peculiar velocity frame.
 Whereas, in the BS model, the
cosmological acceleration dipole, measured by off centre
observers, is emitter's distance dependent and cannot be
eliminated. The same happens for the quadrupole shear term.

However, we do not claim that all the observed matter dipole is
cosmological. In our model, a peculiar velocity dipole induced by
a local inhomogeneity, must be calculated as a local perturbation
of the inhomogeneous background spacetime, i.e. as a local
deviation of the new Hubble law (\ref{8}).

Finally note, that the new Hubble function of this model is not
only cosmic time dependent, as in FLRW models, but observer's
position and angular dependent too. Therefore,
 this may justify
 the difference between its inferred values from observations
performed with different telescopic angles.

\section{Acknowledgments}
This work is partially supported by the spanish Junta de Castilla
y Le\'on projects VA 34/99 and VA 68/00.

\end{document}